\documentclass[
    final            
  ,numberedheadings 
  ]
  {aipproc}

\layoutstyle{6x9}

\begin{document}

\title{Quantum Vacuum Energy in Graphs and Billiards}



\author{L. Kaplan}{
  address={Department of Physics, Tulane University, New Orleans, Louisiana 70118, USA}
}

%

\begin{abstract}
The vacuum (Casimir) energy in quantum field theory is a problem relevant both to new nanotechnology devices and to dark energy in cosmology. The crucial question is the dependence of the energy on the system geometry under study. Despite much progress since the first prediction of the Casimir effect in 1948 and its subsequent experimental verification in simple geometries, even the sign of the force in nontrivial situations is still a matter of controversy. Mathematically, vacuum energy fits squarely into the spectral theory of second-order self-adjoint elliptic linear differential operators. Specifically, one promising approach is based on the small-$t$ asymptotics of the cylinder kernel $e^{-t\sqrt{H}}$, where $H$ is the self-adjoint operator under study. In contrast with the well-studied heat kernel $e^{-tH}$, the cylinder kernel depends in a non-local way on the geometry of the problem. We discuss some results by the Louisiana-Oklahoma-Texas collaboration on vacuum energy in model systems, including quantum graphs and two-dimensional cavities. The results may shed light on general questions, including the relationship between vacuum energy and periodic or closed classical orbits, and the contribution to vacuum energy of boundaries, edges, and corners.\end{abstract}

\maketitle


\section{Introduction}

Since Casimir's famous calculation in 1948 showing an attractive force between parallel conducting plates due to vacuum fluctuations of the electromagnetic field~\cite{casimir}, 
forces associated with vacuum energy in quantum field theory have been studied in a wide variety of contexts~\cite{miltonbook}. These range from the bag model of the nucleon~\cite{bagmodel}, to cosmology~\cite{elizalde}, to stabilization of brane world models~\cite{brane}, and to practical applications in micro- and nano-electromechanical systems~\cite{mems}.

Of course, the calculational details in specific applications will depend on the system dimension, the nature of the relevant quantum fluctuating field (e.g., a vector electromagnetic field), and the detailed boundary conditions (e.g., ones that properly take into account the finite plasma frequency in the electromagnetic case). In the examples considered here, we bypass these application-specific details and instead consider a toy model of a scalar field, usually with Dirichlet or Neumann boundary conditions. As we will see, these simple examples will help to elucidate important general questions concerning Casimir forces that are independent of the specific context. These questions relate to proper regularization and renormalization of the formally infinite vacuum energy, the relation of Casimir forces to periodic and closed classical paths, and the role of boundaries, edges, and corners.

Formally, the vacuum energy of a scalar field is given by ${\hbar \over 2}\sum_n\omega_n$, where $\omega_n$ are the eigenfrequencies, given by solutions of $-\nabla^2 \varphi_n={\omega_n^2 \over c^2} \varphi_n$ with the relevant boundary conditions. In the following, we work in units where $\hbar=c=1$. Of multiple methods of regularizing the infinite vacuum energy (including e.g., dimensional regularization), we focus here on the time-splitting regulator, associated with the cylinder kernel $T_t(x,y)=\langle x
| e^{-\sqrt{-\nabla^2} t}|y\rangle$. Defining
\begin{equation}
E_t=-{1 \over 2} {\partial \over \partial t} {\rm Tr}\, T_t = {1 \over 2} \sum_n \omega_n e^{-\omega_n t} \,,
\end{equation}
the physical vacuum energy is given by taking the limit of $E_t$ as $t \to 0$ if this limit exists (i.e. if the divergent terms can be shown to cancel).

\section{Vacuum Energy in Quantum Graphs}
\label{secgraph}

We begin by applying the above approach to quantum graphs, a class of one-dimensional models that have been widely used as approximations for the free-electron theory of conjugated molecules in chemistry, for quantum wire circuits in nanotechnology, and for photonic crystals in optics.  
More generally, quantum graphs provide a useful testing ground for investigating general
questions about quantum chaos, scattering, and spectral theory. A good discussion may be found in a recent survey by Kuchment~\cite{kuchmentsurvey}.

Mathematically, a quantum graph consists of one-dimensional bonds meeting at vertices, with the scalar field satisfying $-\nabla^2 \varphi_n={\omega_n^2 \over c^2} \varphi_n$ on each bond and prescribed boundary conditions at each vertex. For detailed presentations of the mathematical model, see Refs.~\cite{kottossmil,Kuc}. Vacuum energy in quantum graphs has been studied recently by Berkolaiko, Harrison, and Wilson~\cite{bhw}; here we show some results obtained by Fulling, Kaplan, and Wilson~\cite{fkw}.

\subsection{Pistons in One Dimension}
\label{secline}

\begin{figure}
  \raisebox{0.8in}{\includegraphics[width=0.45\textwidth]{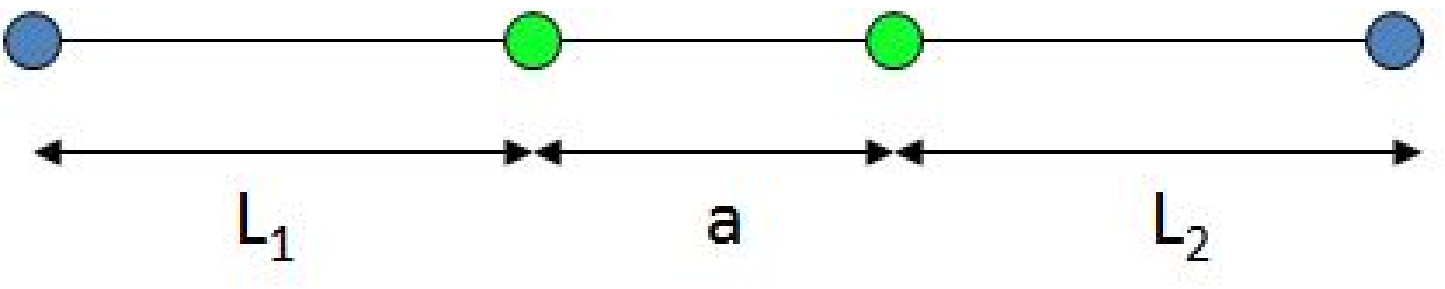}}
  \includegraphics[width=0.45\textwidth]{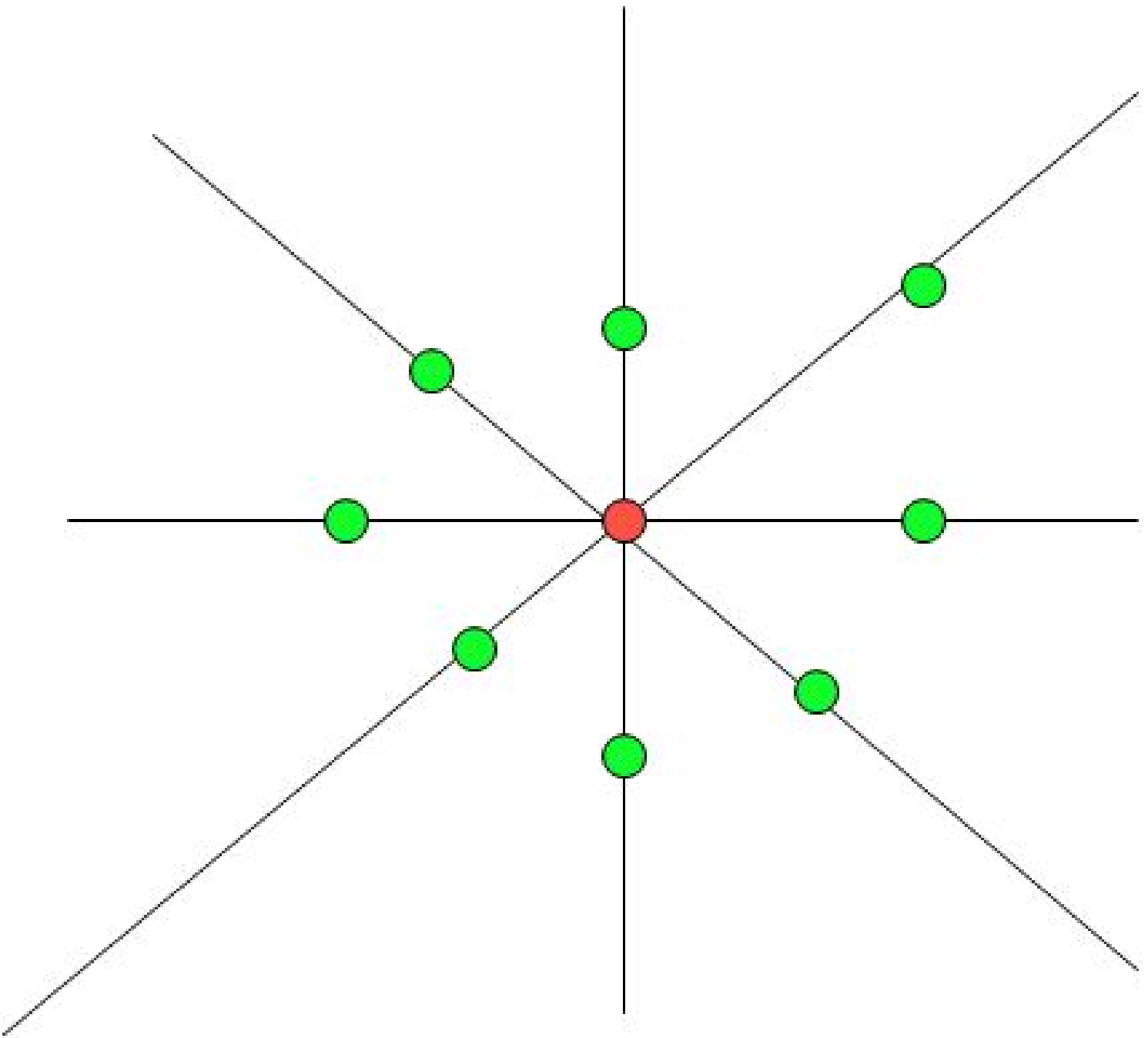}
  \caption{(Left) A line graph consisting of three bonds and four vertices. The two middle vertices are movable pistons. (Right) A star graph with $B=8$ pistons.}
  \label{linegraph}
\end{figure}

The left panel of Figure~\ref{linegraph} shows a simple line graph consisting of three bonds and four vertices. The two middle vertices are to be thought of as movable pistons, i.e. their position may change keeping the total length $L_1+a+L_2$ fixed. The objective is to calculate the vacuum
energy as a function of these positions, and thus to obtain a Casimir force acting on the
pistons~\cite{cavalcanti}. Focusing first on the bond of length $a$ separating the pistons, we note that
the eigenfrequencies are $\omega_n=n \pi /a$, where $n=1, 2, \cdots$ for Dirichlet boundary conditions at the pistons. We then have
\begin{eqnarray}
{\rm Tr} \,T_t&=&  \sum_{n=1}^\infty e^{-\pi nt/a} \nonumber\\
 &=& \frac{e^{-\pi t/a}}{1 -e^{-\pi t/a} }  \\
 &=& \frac a{\pi t} - \frac{1}{2} + 
 \frac{\pi t}{12a} +O(t^2) \nonumber \,,
 \end{eqnarray}
and the regularized vacuum energy is
 \begin{equation}
 E_t = \frac  a{2\pi t^2} -\frac{\pi}{24a} + O(t) \,.
 \label{DDenergy}\end{equation}
 Adding the vacuum energy from the other two segments, we find
\begin{equation}
E_t = \frac  {L_1+a+L_2}{2\pi t^2} -\frac{\pi}{24a} -\frac{\pi}{24L_1} -\frac{\pi}{24L_1}+ O(t) \,.
\end{equation}
Thus the divergent term corresponds to a geometry-independent constant energy density, and
is unobservable since it will not contribute to a force on the piston. After safely discarding this constant energy shift, we
may let $t \to 0$ and $L_{1,2} \to \infty$ 
and obtain the well-known finite, attractive force
 \begin{equation}
 F_{DD} \equiv -\, \frac{\partial E}{\partial a} = -\,\frac{\pi}{24a^2}\,.
 \label{DDforce}  \end{equation}
The same result is obtained if Neumann boundary conditions obtain at each vertex ($\omega_n=n \pi /a$ with $n=0, 1, 2, \cdots$). However, if one piston is Dirichlet and
the other Neumann, the analogous calculation yields a repulsive force
\begin{equation}
 F_{DN}  = +\,\frac{\pi}{48a^2}\,.
\label{DNforce}  \end{equation}

Though simple, the calculations yield little or no insight as to why the force may be attractive in some situations and repulsive in others. To obtain such insight, we turn to an alternative perspective. We first note that
\begin{equation}
{\rm Tr} \, T_t=\int dx \, T_t(x,x) \,.
\end{equation}
Now the free cylinder kernel in one dimension is
\begin{equation}
T^0_t(x,y) = \frac{t}{\pi} \frac{1}{(x-y)^2+t^2} \,.
\end{equation}
Then 
$T_t(x,x)$ in a problem with boundaries is obtainable by the method of images as a sum over periodic and closed orbits:
\begin{equation} T_t(x,x) = 
{\rm Re} \sum_p \frac{t}{\pi}  \frac{A_p}{L_p^2+t^2}+{\rm closed \; orbits}
 \,,\end{equation}
where $p$ labels periodic orbits passing through $x$, $L_p$ is the orbit length, and
 $A_p$ is the product of scattering factors at vertices (e.g., $-1$ at a Dirichlet vertex and
 $+1$ at a Neumann vertex.) The expression arising from closed orbits (i.e. orbits starting
 and ending at $x$ but with opposite momenta) is similar, and is omitted here because after integration over $x$, closed orbits in graphs may be shown to give zero contribution to the total energy. This is not the case in two- or three-dimensional billiards, as discussed in Section~\ref{billiards}.
 
In our case, all periodic orbits in the region between the pistons are repetitions of the primitive periodic orbit of length $2a$. Separating out the zero-length orbit ($r=0$) and taking the trace, we find
\begin{equation}
{\rm Tr} \; T_t=\int dx \, T_t(x,x) = \frac{t}{\pi}\frac{a}{t^2} +
{\rm Re} \sum_{r=1}^\infty \frac{t}{\pi}  \frac{4a A^r}{(2ra)^2+t^2}
 \,,
\end{equation}
where $r$ labels the repetition number and $A$ is the product of scattering factors for the primitive two-bounce orbit. 

The asymptotic $t \to 0$ behavior may now be analyzed term by term. We immediately see that all orbits of nonzero length make contributions $\sim t$ to the cylinder kernel, and thus 
finite $t-$independent contributions to the energy. The $t \to 0$ divergence is seen to be associated exclusively with the zero-length periodic orbits, which exist at every point $x$ and yield the divergent but constant and geometry-independent energy density.

The periodic orbit sum converges. Differentiating ${\rm Tr} \; T_t$ with respect to $t$ to obtain the vacuum energy and then with respect to $a$ to obtain the force on a piston, we have
\begin{equation}
F=
- {1 \over 4 \pi a^2}\sum_{r=1}^\infty \frac{A^r}{r^2} \,.
\label{fper}
\end{equation}
We note that $A=+1$ if the pistons are both Neumann or both Dirichlet, and $A=-1$ for mixed boundary conditions. Eq.~(\ref{fper}) thus reproduces the  results of Eqs.~(\ref{DDforce}) and (\ref{DNforce}); moreover we see that the sign of the force arises from reflection phases, and is already correctly obtained if we consider only the phase associated with the shortest periodic orbit ($r=1$).

\subsection{Pistons in General Star Graphs}
\label{secstar}

We turn our attention to star graphs, an example of which is pictured in the right panel of Fig.~\ref{linegraph}. A total of $B$ line segments of large length $L$ meet at the central vertex, where the wave function satisfies Kirchhoff boundary conditions: (i) continuity $\varphi_j(0)=\varphi_k(0)$ for all $j, k=1 \cdots B$ and (ii) current conservation $\sum_{j=1}^B \varphi_j'(0)=0$, where $\varphi_j'(0)$ is the outward derivative along bond $j$. Along each segment $j$, a piston
is located at distance $a_j$ from the central vertex, and the boundary condition imposed by the piston may be Dirichlet (reflection with phase $-1$), Neumann (reflection with phase $+1$), or reflection with an arbitrary phase $e^{i\theta_j}$ (to break time reversal symmetry). We will be interested in computing the dependence of the energy on the piston positions, i.e. in the Casimir force on the pistons.

We focus initially on the ``interior'' of our system, i.e. on the graph consisting of $B$ bonds of length $a_1 \cdots a_B$, and excluding the space beyond the pistons. For $B>2$ and generic $a_j$, no analytic expression exists for the spectrum or for the vacuum energy, and a numerical approach must be employed. For a general quantum graph, the
eigenfrequencies are given by solutions of a characteristic equation ${\rm det} \; h(\omega)=0$~\cite{kottossmil}; in the case of a star graph with irrationally related bond lengths $a_j$ the relevant equation becomes 
\begin{equation}
\sum_{j=1}^B \tan(\omega a_j + \theta_j) = 0 \,,
\end{equation}
where $\theta_j=0$ or $\pi$ for a Neumann or Dirichlet piston on 
bond~$j$, respectively~\cite{fkw}. If we numerically obtain in this way all eigenfrequencies $\omega_n$ up to a cutoff $\omega_{\rm max}$, we may write
\begin{equation}
E^{\rm num}_t=\frac{1}{2} \sum_{\omega_n <\omega_{\rm max}} \omega_n e^{-\omega_n t}=E_t+O(e^{-\omega_{\rm max}t}) \,.
\label{numer}
\end{equation}
Since the error associated with omitting eigenfrequencies $\omega_n >\omega_{\rm max}$ is 
$O(e^{-\omega_{\rm max}t})$, we must consider $\omega_{\rm max} t \gg 1$.

Now we turn to the ``outside,'' i.e. the segments $a_j \le x_j \le L$ located beyond the pistons. From Section~\ref{secline} we know that the outside energy consists of finite terms that decay as $1/L$ and may therefore be neglected for large $L$, plus a divergent $1/t^2$ term associated with a geometry-independent constant energy density. The divergent terms, as before, will combine with the divergent part of the interior vacuum energy to yield a constant energy shift
$BL/2\pi t^2$ proportional to the total length $BL$ and independent of the piston positions. To obtain the physical forces on the pistons for large $L$ we thus need only
to subtract from the interior energy the divergence proportional to the total interior length $\sum_j a_j$. The physically observable energy is then given by
\begin{equation}
E^{\rm finite}_t= E^{\rm num}_t -E_t^{\rm Weyl}\,,
\end{equation}
where the divergence coming from integrating the Weyl density in one dimension $\rho(\omega)=\sum_j a_j/\pi$ between $0$ and $\omega_{\rm max}$ is
\begin{equation}
E_t^{\rm Weyl} = \left(\sum_{j=1}^B a_j\right) \cdot \frac{ [1-(\omega_{\rm max}t+1)e^{-\omega_{\rm max}t}]}{2 \pi t^2} \,.
\label{weyl}
\end{equation}
Expressing the 
finite part of $E_t$
as a power series,
\begin{equation}
E^{\rm finite}_t =E_0 + \alpha_1 t +\alpha_2 t^2 +\cdots \,
\end{equation}
and numerically evaluating $E^{\rm finite}_t$ for several values of the cutoff $t$ with
$\omega_{\rm max}^{-1} \ll t \ll {\rm min} (a_j)$, we easily obtain the vacuum energy $E_0$
for any given star graph to any desired order of accuracy.

\begin{figure}
  \raisebox{0.0in}{\includegraphics[width=0.45\textwidth]{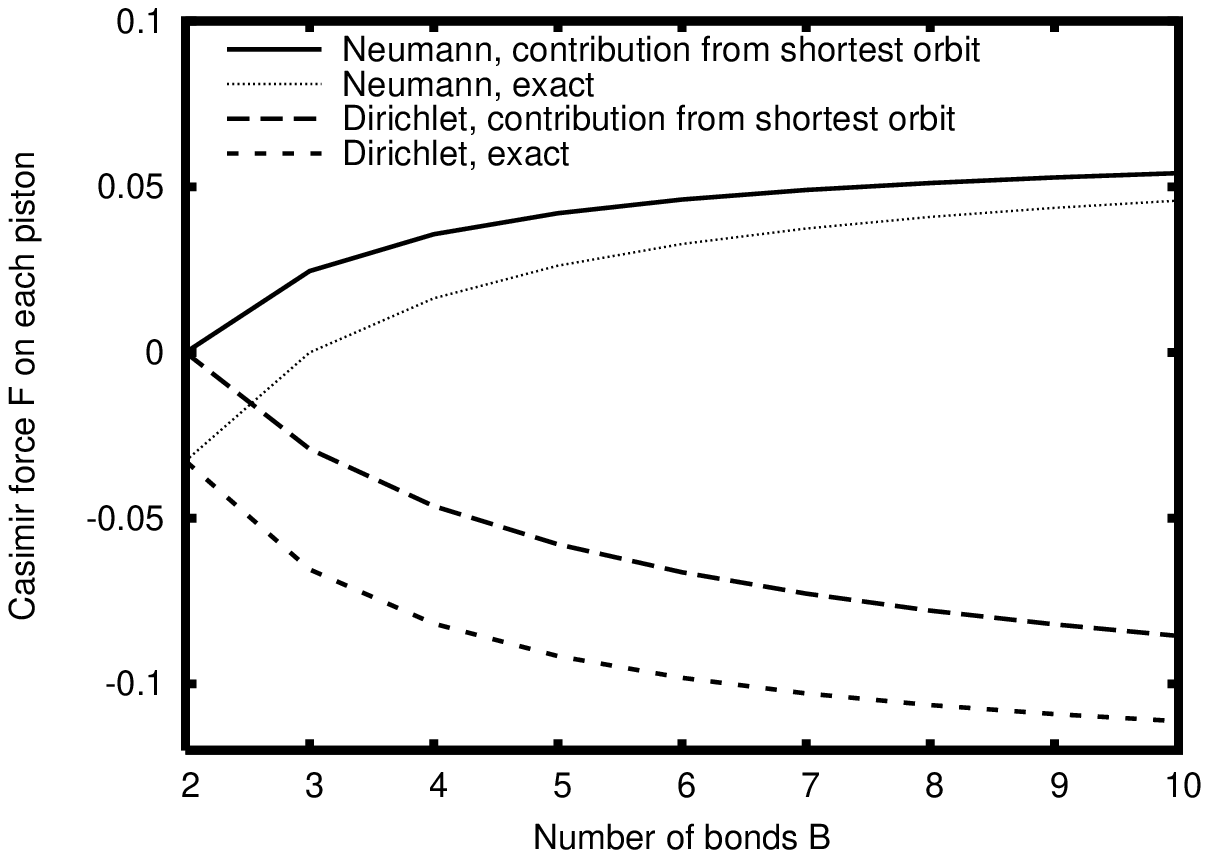}}
  \includegraphics[width=0.45\textwidth]{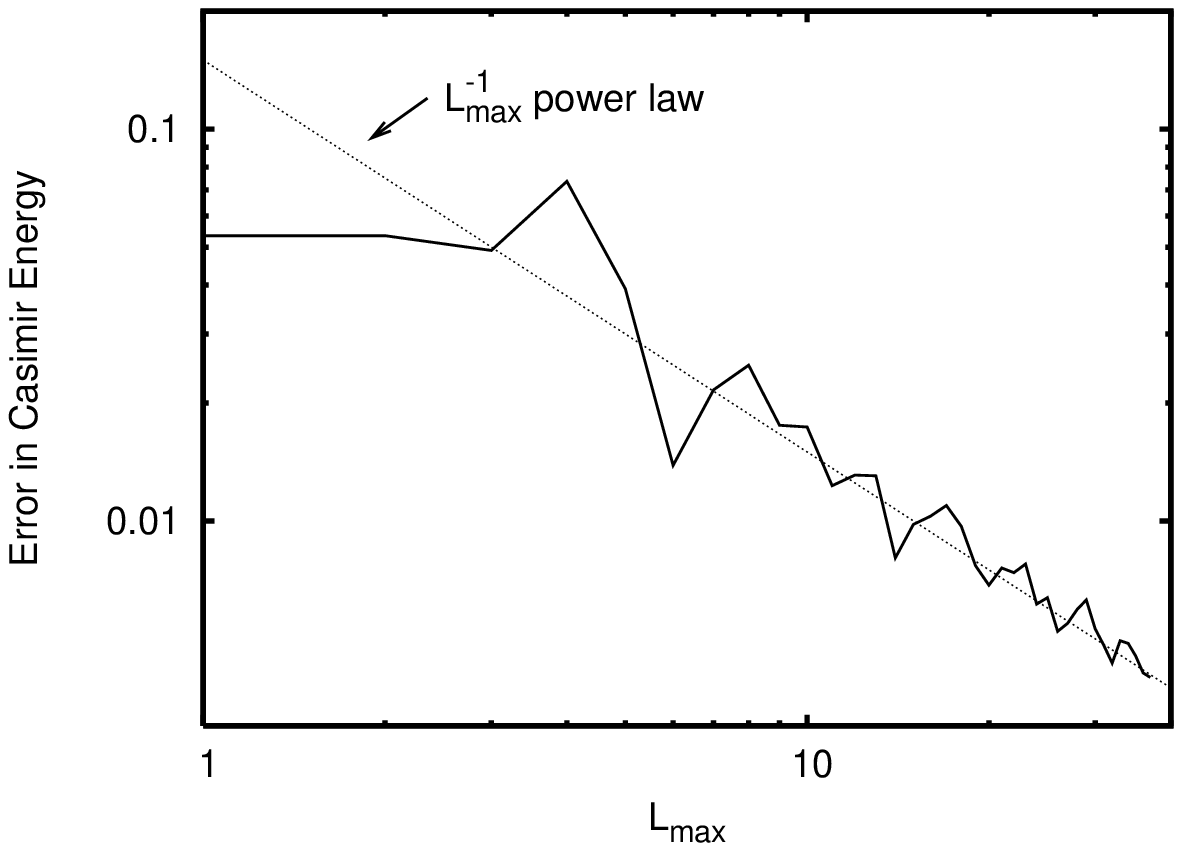}
  \caption{(Left) The force on a piston in a star graph with $B$ bonds
of length $1$ and either
Neumann or Dirichlet boundary conditions at each piston is computed
using only repetitions of the shortest periodic orbit and
compared with the exact answer. Positive values indicate repulsive forces. (Right) The difference between the exact Casimir energy
$E_0$ and a periodic orbit sum including all orbits up to length $L_{\rm max}$ is shown for a star graph with four bonds of length $1.1$, $1.6176$, $1.2985$, and $1.1159$, and a
Neumann piston at the end of each bond~\cite{fkw}.}
\label{figstarresults}
\end{figure}

To attain a more physical understanding, we wish to compare the numerical results with periodic orbit sums. For example, taking all pistons to have Neumann boundary conditions and summing only over repetitions of the shortest periodic orbits (i.e., over orbits bouncing back and forth in a single bond), we obtain
\begin{equation}
E_N^{\rm shortest} = \frac{\pi}{48} \left(1-\frac{24 \ln 2}{\pi^2B}
+\cdots\right) \sum_{j=1}^B \frac{1}{a_j} \,,
\end{equation}
which for large $B$ compares well to  the analytic result
$\frac{\pi}{48}\left(1-\frac{3}{B}\right)\frac{B}{a}$ for
$B$ {\it equal-length} bonds. Similar results are obtained in the Dirichlet case.
As illustrated in the left panel of Figure~\ref{figstarresults}, both repulsive behavior in the Neumann
case and the attractive behavior in the Dirichlet case are well
explained by considering only the shortest periodic orbit. For a better quantitative
approximation we may add contributions from orbits of longer length. Figure~\ref{figstarresults} (right panel) shows the convergence to the numerically exact Caimir energy $E_0$ for a typical star graph with Neumann pistols. Here the error scales
as $L_{\rm max}^{-1}$, where $L_{\rm max}$ is the length of the longest orbit included. For mixed Dirichlet or Neummann boundaries, or for arbitrary phases at the pistons, the convergence is shown to be $L_{\rm max}^{-3/2}$~\cite{fkw}.

\section{Vacuum Energy in Rectangles, Pistons, and Pistols}
\label{billiards}

We now extend the approach of Section~\ref{secgraph} to two-dimensional billiards (the extension to the three-dimensional case and the electromagnetic field is also straightforward~\cite{liuthesis}.) An important motivation for this work~\cite{rect} is to investigate the physical reality of the outward force on the walls of a square or cubic box, as obtained by Lukozs using naive renormalization (i.e., by simply discarding infinite terms) and ignoring the outside of the box~\cite{lukosz}. 

We begin with a rectangle of sides $a$ and $b$. As for a line segment (Section~\ref{secline}), we can use the method of images to evaluate ${\rm Tr}\; T_t$, and thus the regularized vacuum energy $E_t$. Each term in the image sum may be associated with a classical path leading from $x$ to $x$ in the rectangular, and these terms may be classified by the number of bounces the path makes off the vertical and horizontal walls. Periodic paths make an even number of bounces $2j$ off the vertical sides and an even number of bounces off the horizontal sides. The resulting contribution to the vacuum energy is
\begin{eqnarray} E_{t,{\rm Periodic}} &=& \frac{ab}{2\pi t^3}
-\frac{ab}{2\pi} \sum_{k=1}^\infty (-1)^{\eta_{0k}} \frac{(2kb)^2 -2t^2}
 {[t^2 +(2kb)^2]^{5/2}} 
 -\frac{ab}{2\pi} \sum_{j=1}^\infty (-1)^{\eta_{j0}} \frac{(2ja)^2 -2t^2}
 {[t^2 +(2ja)^2]^{5/2}}  \nonumber \\
&& - \frac{ab}{\pi} \sum_{j=1}^\infty\sum_{k=1}^\infty (-1)^{\eta_{jk}}
 \frac{(2ja)^2 +(2kb)^2 -2t^2}{[t^2 +(2ja)^2 +(2kb)^2]^{5/2}} \,,
 \end{eqnarray}
where $\eta_{jk}$ is the number of Dirichlet bounces for a given orbit, and we have separated out the purely vertical and purely horizontal periodic  orbits ($j=0$ and $k=0$, respectively), as well as the zero-length orbit $j=k=0$. As in the one-dimensional case, the zero-length orbit contributes a divergent but constant  and geometry-independent energy density, i.e., a divergence proportional to the total area $ab$. Assuming all Neumann or
all Dirichlet sides, so that all $\eta_{jk}=0$, we have
\begin{equation} E_{t,{\rm Periodic}} = \frac{ab}{2\pi t^3}
-\,\frac{\zeta(3)}{16\pi}\left({a \over b^2}+{b \over a^2}\right) 
-\,\frac{ab}{8\pi} \sum_{j=1}^\infty\sum_{k=1}^\infty
 \left(a^2j^2 + b^2k^2\right)^{-3/2} +O(t^2)\,. 
 \end{equation}
 In contrast with the one-dimensional case, here the non-periodic closed orbits (ones that
 make an odd number of bounces off the horizontal sides, or an odd number of bounces off the 
vertical sides, or both), make a nonzero contribution to the total energy, including a divergent contribution proportional to the system perimeter. Combining periodic and non-periodic terms we obtain
 \begin{eqnarray} E_{t} &=& \frac{\rm Area}{2\pi t^3} \mp\, \frac {\rm Perimeter}{8\pi t^2} 
-\,\frac{\zeta(3)}{16\pi}\left({a \over b^2}+{b \over a^2}\right) \nonumber \\
&-&\,\frac{ab}{8\pi} \sum_{j,k=1}^\infty
 \left(a^2j^2 + b^2k^2\right)^{-3/2} 
  \pm{\pi \over 48}\left( {1 \over a}+{1 \over b}\right) +O(t^2)  \,,
  \label{rectenergy}
 \end{eqnarray} 
 where $\mp$ refers to Neumann or Dirichlet boundaries, respectively. Naively
 discarding the divergent terms and differentiating with respect to $a$ we find
an attractive force for $a \ll b$ (as expected in the limit of two infinite parallel plates), but a repulsive force for the square $a=b$.

\begin{figure}
  \raisebox{0.2in}{\includegraphics[width=0.45\textwidth]{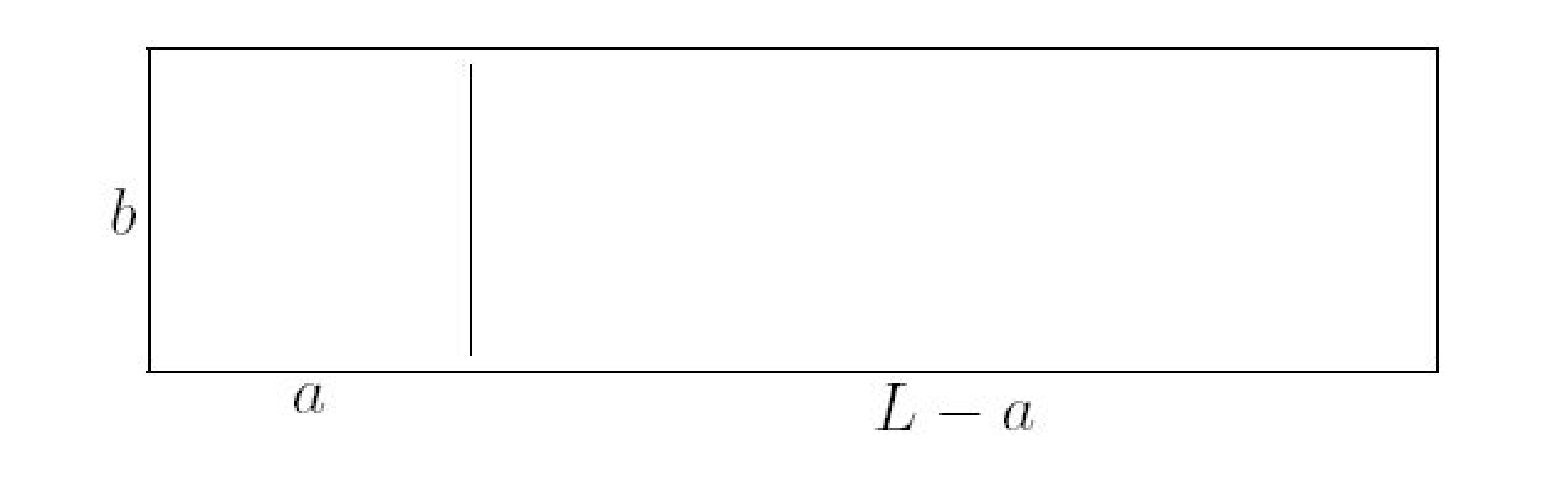}}
  \includegraphics[width=0.45\textwidth]{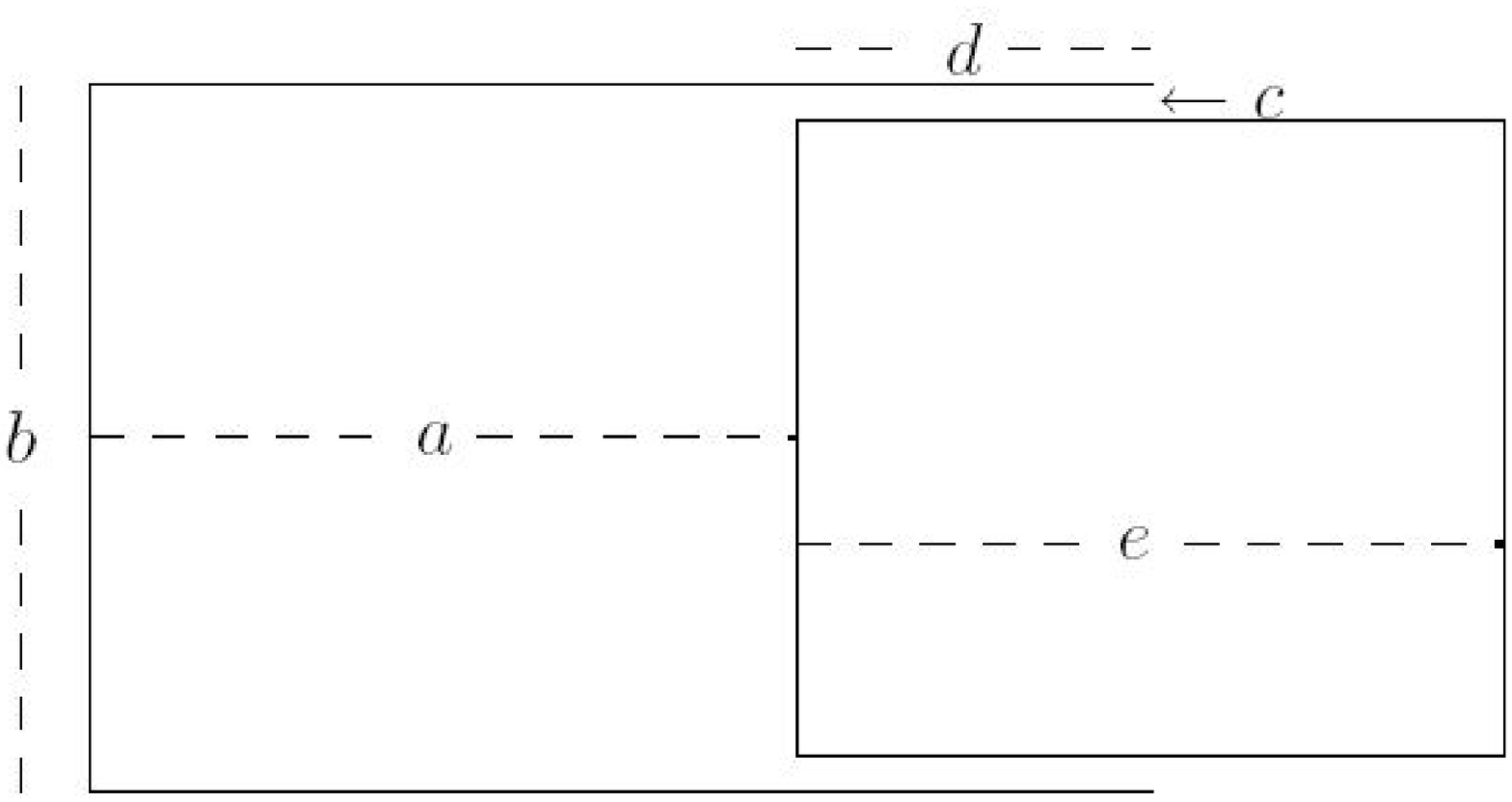}
  \caption{(Left) A piston in a rectangular box. (Right) A pistol configuration, consisting
  of a ``barrel'' and a ``bullet.''}
\label{figrect}
\end{figure}

The above analysis has two (related) problems: discarding divergent terms and ignoring the outside of the box. Both may be cured by considering a piston configuration~\cite{cavalcanti}, as discussed in Section~\ref{secgraph} and illustrated in Figure~\ref{figrect} (Left). Adding contributions from the $a \times b$ rectangle and the $(L-a) \times b$ rectangle, we see that the divergent terms combine to yield contributions proportional to the total system area, and total boundary length, and thus independent of the position of the piston. Other than these divergent terms, the only contribution from the
$(L-a) \times b$ rectangle that survives the $L \to \infty$ limit is
$\frac{\zeta(3)}{16\pi}\left({L-a \over b^2}\right)$; combining this term with the finite
part of Eq.~(\ref{rectenergy}) and differentiating with respect to $a$ we obtain a finite
Casimir force,\
\begin{equation}
F_{\rm piston}=
  \frac{\pi}{b^2} \sum_{j,k=1}^\infty k^2
  K_1'\left(2\pi jk\,\frac ab \right) \,.
  \end{equation}
This force is always attractive (decaying exponentially for $a \gg b$ and reducing to the parallel plate limit for $a \ll b$).

Finally, to approach the original motivating situation of a box with a loose lid~\cite{lukosz} and to address the question of what happens when the external shaft is not present, we consider the ``pistol'' configuration illustrated in the Right panel of Figure~\ref{figrect}. Here  all system dimensions other than possibly the gap $c$ are assumed to be large compared to the ultraviolet cutoff $t$. We then use scaled variables
$c = rt$, $a = st$, $b=ut$, $d =(\ell-s)t$, where $s, u, \ell \gg 1$, and for all Dirichlet boundaries obtain
\begin{eqnarray} E_{\rm pistol}&=& 
\frac{us}{\pi t}\sum_{k=1}^\infty 
 \frac{1-2k^2u^2}{(1+4k^2u^2)^{5/2}} 
+ \frac{us}{\pi t}\sum_{j=1}^\infty 
 \frac{1-2j^2s^2}{(1+4j^2s^2)^{5/2}}
  \nonumber \\
&&{}+\frac{2us}{\pi t} \sum_{j=1}^\infty\sum_{k=1}^\infty
 \frac{1 - 2j^2s^2 - 2k^2u^2 }{ (1+ 4j^2s^2 + 4 k^2 u^2)^{5/2}} 
\nonumber \\
&&{}+ \frac s{2\pi t}   \sum_{j=1}^\infty 
\frac{-1+4j^2s^2}{(1+4j^2s^2)^2 } 
+\frac{2r(\ell-s)} {\pi t} \sum_{k=1}^\infty
 \frac{1 - 2k^2r^2}{(1 + 4k^2r^2)^{5/2} } 
\end{eqnarray}
In the case of a narrow chamber, $a \ll b^{1/3}c^{2/3}$, we recover an attracive force
$\sim 1/a^2$, as required in the parallel plate limit. In the opposite case of a long
chamber, $a \gg b^{1/3}c^{2/3}$, we find that the gaps of width $c$ dominate and we obtain an $a$-independent force that is attractive for $c> \alpha t$ and repulsive for $x<\alpha t$, where $\alpha \approx 0.5888$. This last situation, however, is least convincing physically, as we need to be in a regime where the gap dimension is smaller than the cutoff. See Ref.~\cite{rect} for a detailed discussion.
  
\section{General Billiards}

We note that the numerical approach to calculating the vacuum self-energy, applied to general quantum star graphs in Section~\ref{secstar}, may be equally well applied to two- or three-dimensional systems, provided the spectrum may accurately be computed numerically up to some maximum frequency $\omega_{\rm max}$. Of course the appropriate Weyl term
containing all $t \to 0$ divergences must be subtracted from the numerical sum (\ref{numer}) before the
numerical limit $t \ll 1$ may be considered. For example in the case of the interior of a three-dimensional cylinder with polygonal cross sections and Dirichlet boundary conditions~\cite{abalo}, Eq.~(\ref{weyl})
becomes
\begin{equation}
E_t^{\rm Weyl}(t)=\frac{1}{2}  
\int_{-\infty}^{\infty} \frac{dk}{2\pi}\,
\int_0^{\omega_{\rm max}} d\omega \sqrt{k^2+ \omega^{2}} \,e^{-t \sqrt{k^2+ \omega^{2}}}
\left(\frac{\gamma\;{\rm Area}}{2\pi}-\frac{{\rm Perimeter}}{4\pi}\right)+\frac{C}{48\pi t^2} \,,
\label{e-div}
\end{equation}
where the area and the perimeter refer to the polygonal cross section, $C=\sum_i \left(\frac\pi{\alpha_i}-\frac{\alpha_i}\pi\right)$ with $\alpha_i$ the interior corner angles of the polygon, and $\omega_{\rm max}$ is the maximal eigenvalue obtained numerically for the polygon. An additional divergent term $\ln t$ must be considered in the presence of boundary curvature. These approaches are now being applied to study the self-energy of stadium-shaped and elliptical cavities, as well as to investigate the self-energy associated with the {\it outside} of a billiard of arbitrary shape.

\section{Conclusions}

We have seen that careful regularization and renormalization (including both inside and outside contributions) are needed to obtain physically meaningful vacuum energies and Casimir forces. Classical orbit approaches, including both periodic and non-peridoic orbits, produce exact results in simple cases and may allow for good approximations where exact solutions are nonexistent, including general quantum graphs and polygonal billiards. Furthermore, such semiclassical approximations may be compared with results obtained
numerically by directly summing eigenfrequencies and subtracting known divergences associated with the Weyl part of the spectrum.
Intelligent combination of analytical and numerical tools can be a promising tool for furthering our understanding of Casimir forces in general geometries.


\begin{theacknowledgments}
  This work was supported in part by the NSF under Grant No. PHY-0545390.
\end{theacknowledgments}

\bibliographystyle{aipproc}   

\end{document}